\pdfoutput=1
\documentclass[aps,12pt,superscriptaddress]{revtex4-2}
\usepackage{bm}
\usepackage{mathrsfs}
\usepackage{amsmath,amsfonts}
\usepackage{wasysym}
\usepackage{comment}
\usepackage[colorlinks=true,citecolor=blue,urlcolor=blue,linkcolor=black,pdfstartview=FitH,bookmarksopen]{hyperref}

\usepackage{color,graphicx}
\graphicspath{{Images/}}
\usepackage{subfigure}

\newcommand{\ii}{\mathrm{i}}
\newcommand{\cC}{ {\cal C} }
\renewcommand\d{\partial}
\newcommand\vareps{\varepsilon}
\newcommand\E{\mathbf{E}}
\newcommand\x{\mathbf{x}}

\renewcommand\d\partial
\renewcommand\u{\mathbf{u}}
\renewcommand\l{\mathbf{l}}
\newcommand\<\langle
\renewcommand\>\rangle
\newcommand\+\dagger
\newcommand\A{\mathbf{A}}
\newcommand\B{\mathbf{B}}
\newcommand\bnabla{\bm{\nabla}}
\renewcommand\a{\mathbf{a}}
\renewcommand\b{\mathbf{b}}

\newcommand\s{\mathbf{s}}

\newcommand{\beq}{\begin{equation}}
\newcommand{\eeq}{\end{equation}}
\newcommand{\beqn}{\begin{eqnarray}}
\newcommand{\eeqn}{\end{eqnarray}}

\begin{document}

\date{September 2023}
\title{Nonlinear Lifshitz Photon Theory in Condensed Matter Systems}
\author{Yi-Hsien Du}
\affiliation{Kadanoff Center for Theoretical Physics, University of Chicago, Chicago, Illinois 60637, USA}
\affiliation{Kavli Institute for Theoretical Physics, University of California, Santa Barbara, California 93106, USA}

\author{Cenke Xu}
\affiliation{Department of Physics, University of California, Santa Barbara, California 93106, USA}

\author{Dam Thanh Son}
\affiliation{Kadanoff Center for Theoretical Physics, University of Chicago, Chicago, Illinois 60637, USA}

\begin{abstract}
We present an interacting theory of a $U(1)$ gauge boson with a quadratic dispersion relation, which we call the ``nonlinear Lifshitz photon theory.''  The Lifshitz photon is a three-dimensional generalization of the Tkachenko mode in rotating superfluids. Starting from the Wigner crystal of charged particles coupled to a dynamical $U(1)$ gauge field, after integrating out gapped degrees of freedom, we arrive at the Lagrangian for the nonlinear Lifshitz photon. The symmetries of the theory include a global $U(1)$ 1-form symmetry and nonlinearly realized ``magnetic" translation and rotation symmetries. The interaction terms in the theory lead to the decay of the Lifshitz photon, the rate of which we estimate. We show that the Wilson loop, which plays the role of the order parameter of the spontaneous breaking of the 1-form global symmetry, deviates from the perimeter law by an additional logarithmic factor. We explore potential connections to other condensed matter systems, with a particular focus on quantum spin ice and ferromagnets. Finally, we generalize our theory to higher dimensions.
\end{abstract}

\maketitle

\tableofcontents
\section{Introduction}

Recent years have seen a significant surge of interest in exotic quantum field theories~\cite{Pretko:2016kxt,Pretko:2016lgv,Pretko:2017kvd,Seiberg:2020bhn,Seiberg:2020wsg,Gorantla:2020xap,PhysRevB.106.045112,Gorantla_2023}.  These field theories frequently have unusual symmetries that are not possible in ordinary Lorentz-invariant theories. One example that exhibits many features typical of these theories is the so-called ``Lifshitz scalar'' theory, whose Lagrangian contains a term involving four spatial derivatives of the field, while the usual term with two spatial derivative is absent. These theories manifest exotic symmetries that limit the mobility of a single charge (the ``dipole symmetry'' or, in some cases, its generalization called ``higher multipole symmetries''). Recently, a nonlinear Lifshitz scalar theory, characterized by nonlinear dipole and multipole symmetries, has been found to describe the Tkachenko mode of a rotating two-dimensional superfluids~\cite{du2022noncommutative}.  In this system, the dipole symmetry is simply the symmetry of magnetic translations, while the higher multipole symmetry is that of the magnetic rotation.  Since these are exact symmetries of the system at the microscopic level, the nonlinear quantum Lifshitz theory of the Tkachenko mode does not require fine tuning~\cite{Lake_2022}.  The theory of the Tkachenko mode allows a convenient formulation as a noncommutative field theory~\cite{Douglas:2001ba,Rubakov-NC,du2021noncommutative}.  The form of the cubic self-interaction of the Tkachenko scalar is fixed by the symmetries and determines the decay rate of the Tkachenko mode.

In this paper, we consider three-dimensional analogs of the Tkachenko mode in two dimensions.  We show that such an extension is indeed possible, and the resulting theory describes a $U(1)$ gauge boson with a quadratic dispersion relation---the ``Lifshitz photon.''  This photon can be interpreted as the Nambu-Goldstone boson of the spontaneously broken one-form symmetry.

The paper is organized as follows.  We first analyze in detail in Sec.~\ref{Lifshitz} a prototypical example---a Wigner crystal of charged particles, immersed in a fixed uniformed compensating background, interacting with a dynamical $U(1)$ gauge field.  We show that the low-energy mode in this model is a Lifshitz photon.  We then (in Sec.~\ref{Nonlinear}) show that the model exhibits a nonlinear version of the dipole symmetry, which can be used to constrain the form of the self-interaction of the Lifshitz photon, and which allows us to determine the decay rate of the latter.  In Sec.~\ref{sec:prospects} we comment on one peculiar feature of the higher-form symmetry in our system: a logarithmic correction to the perimeter law for the 't Hooft loop.  In Sec.~\ref{sec:lattice} we discuss some spin systems where the Lifshitz photon may be realized, and in Sec.~\ref{sec:ferro} we speculate on a possible deconfined phase of ferromagnetism in which the symmetries of the ferromagnetic nonlinear sigma model are realized through the Lifshitz photon.

\section{Prototype: Wigner crystal coupled to dynamical $U(1)$ gauge field}\label{Lifshitz}

The prototypical system that we will consider is the system of charged
particles (``ions'') forming a lattice, interacting with a $U(1)$ gauge field $A_\mu$.  At this moment we do not specify the nature of $A_\mu$; it can be the physical electromagnetic field or an emergent $U(1)$ gauge field of, e.g., a three-dimensional spin liquid.  We assume that there is a neutralizing background with charge of the opposite sign, and that this background has no dynamics of its own.

In this Section we will construct a linear theory of excitations and
show that the lowest mode is a ``Lifshitz photon,'' i.e., a
quasiparticle with two transverse polarizations and a quadratic
dispersion.  One can describe the lattice of the ions in terms of the
displacement field of $\u(\x)$. Let $n_0$ be the equilibrium density
of the ions, $M$ and $e$ be the mass and the charge of the ion.  The
Lagrangian of the system is
\begin{equation}
  \mathcal L = \frac{Mn_0}2 \dot\u^2 - 
  \frac\mu 4
  \left(\d_i u_j + \d_j u_i - \frac23\delta_{ij}\d_k u_k\right)^2
  - \frac K2 (\d_i u_i)^2
  + e n_0 \u\cdot \E + \frac{\E^2}{2} - \frac{\B^2}{2} \, , \label{laga}
\end{equation}
where $\E = \bnabla A_0 - \d_t\A$, $\B=\bnabla\times\A$, and $\mu$ and $K$ are the shear and bulk moduli of the lattice (for
simplicity we assume rotational invariance).  We will find later that
our mode has quadratic dispersion $\omega \sim q^2 \ll q$, so at low momenta
the terms $\dot u^2$ and $E^2$ can be omitted since $\dot u^2 \ll(\d
u)^2$, and $E^2\ll B^2$. The Lagrangian now becomes
\begin{equation}\label{L-simplified}
  \mathcal L = 
  -\frac\mu 4\left(\d_i u_j + \d_j u_i - \frac23\delta_{ij}\d_k u_k\right)^2
  - \frac K2 (\d_i u_i)^2
  + e n_0 \u\cdot \E  - \frac{\B^2}{2} \, .
\end{equation}
Variation of the action with respect to $A_0$ gives rise to condition
\begin{equation}
  \bnabla \cdot \u = 0 \, ,
\end{equation}
which means that the displacement is constrained to be transverse.  
This condition can be solved:
\begin{equation}\label{u-a}
  \u = \frac1{g n_0} \bnabla \times \a \, ,
\end{equation}
where for future convenience we have introduced a for now unspecified
constant $g$.  We will now integrate out $A_i$.  There are two terms
in the Lagrangian that contain $A_i$,
\begin{equation}
  - \frac eg (\bnabla\times\a) \cdot \d_t \A
  - \frac{\B^2}{2} = \frac eg \d_t \a \cdot \B - \frac{\B^2}{2} +
  \text{total derivative} \, .
\end{equation}
It is tempting to integrate over $\B$ here to obtain $\dot\a^2$, but
that would be incorrect: the three components of $\B$ are not
independent but satisfies the condition $\bnabla\cdot\B=0$.  We
introduce a Lagrange multiplier enforcing this constraint:
\begin{equation}\label{Lagrange-mult}
  \frac eg \d_t \mathbf a \cdot \B - \frac{\B^2}{2}
  + \frac eg a_0 (\bnabla\cdot\B) \, .
\end{equation}
Now we can integrate over $\B$.  Setting $\B$ at the saddle point located at
\begin{equation}\label{B-e}
  \B = \frac eg (\d_t \mathbf a - \bnabla a_0) \, .
\end{equation}
we get 
\begin{equation}
  \frac{e^2}{2g^2} (\bnabla a_0-\d_t\a)^2 \, .
\end{equation}
Together with the elastic energy, the Lagrangian is
\begin{equation}\label{Lifshitz_photon_theory}
  \mathcal L = \frac{e^2}{2g^2} e_i e_i - \frac\mu{2g^2n_0^2}(\d_i b_j)^2 \, ,
\end{equation}
where $e_i=\d_i a_0-\d_t a_i$, and $b_i=\epsilon_{ijk}\d_j a_k$.  The
theory now has a $U(1)$ gauge invariance under $a_\mu\to
a_\mu+\d_\mu\lambda$; the excitation of the model is a ``Lifshitz
photon'' with the quadratic dispersion relation
\begin{equation}
  \omega = \frac{\sqrt \mu}{n_0} q^2 \, .
\end{equation}

That the dispersion relation is quadratic is due to the absence of the
usual term $\mathbf b^2$ from the Lagrangian, which can be traced back
to the translational invariance. Under translation the lattice
displacement is shifted by the constant vector: $\u \to \u + \mathbf
c$.  Taking into account Eq.~(\ref{u-a}) this is implies a shift
symmetry acting on $\mathbf b$, which forbids the term $\mathbf b^2$
in the Lagrangian.

The photon here is a dual to the physical photon.  That can be seen
from Eq.~(\ref{B-e}), which equates, up to a proportionality
coefficient, the Lifshitz electric field with the physical magnetic
field.  One can also see this duality by introducing an external
electric charge $e$ located at $\x=\mathbf 0$ into the system.  This
is done by adding to Eq.~(\ref{L-simplified}) a term
$e\delta(\x)A_0(\x)$.  Then variation over $A_0$ gives rise to the
equation
\begin{equation}
  en_0 \bnabla \cdot \u = e\delta(\x) \Rightarrow
  \bnabla \cdot \b = g\delta(\x) \, ,
\end{equation}
which means that electric charge is a magnetic monopole in the dual
theory, with the previously introduced parameter $g$ playing the role
of the monopole charge.

Vice versa, if we have a magnetic monopole of charge $g$ then we
have to modified the last term in Eq.~(\ref{Lagrange-mult}) as
\begin{equation}
   \frac eg a_0 (\bnabla\cdot\B - g\delta(\x)) \, .
\end{equation}
That means introduction of a point-like electric charge $-e$ in
Lifshitz photon theory.


The Lifshitz photon can be interpreted as the result of a hybridization between two vector modes: the transverse phonon, carried by the transverse components of the displacement $\mathbf{u}$, and the photon $A_\mu$.  The mixing between the two mode is due to the dipole coupling $\mathbf{u}\cdot\mathbf{E}$.

Let us also note here that the situation described in our model should occur exactly at the quantum critical point of the ferroelectric phase transition~\cite{Kittel}.  Such a phase transition can be described by the Lagrangian~(\ref{laga}) where $\mathbf{u}$ is interpreted as the electric polarization density of the material, with the addition a potential energy term,
\begin{equation}
  \mathcal L_{\rm pot} = - \frac{Mn_0}2 \omega_0^2 \mathbf{u}^2 - O(u^4).
\end{equation}
Here $\omega_0$ is the frequency of the transverse optical phonon (at zero wavenumber).  It is known that the dielectric constant diverges as one approaches the critical point, $\omega_0\to 0$, hence the velocity of light in the material goes to zero.  Exactly at the phase transition point, $\omega_0=0$, the photon velocity vanishes, and hence it is natural for the dispersion relation for the photon to become quadratic.

Note that the quadratically dispersing photon can only be seen very close to the ferroelectric phase transition, i.e., when the photon velocity is smaller than the phonon velocity.  This requires the dielectric constant larger than $10^{10}$.  To compare, the largest value of the dielectric constant that has been reported in isotope-substituted SrTiO$_3$ is $\lesssim 2\times 10^5$~\cite{Itoh:2000}.

\section{Nonlinear Lifshitz photon theory}\label{Nonlinear}

In the previous Section, we have presented a model in three
spatial dimensions that reduces to the Lifshitz photon theory at low
energy in Eq.~(\ref{Lifshitz_photon_theory}). The model is a Wigner
crystal coupled to a dynamical $U(1)$ gauge field.  We now investigate
this model at the nonlinear level.  We will find that the model
possesses a nonlinear version of the dipole symmetry as well as a
nonlinear multipole symmetry.  Our construction can be considered as a
three-dimensional generalization of the theory of the Tkachenko mode
constructed in Ref.~\cite{du2022noncommutative}.

A state of a solid can be described by a map between the external
coordinate $x^a$ and the coordinate frozen into the solid $X^a(t,
x^i)$, where $a=1,2,3$ in three spatial dimensions. The lattice
displacement is $u^a = \delta^a_i x^i -X^a$. The particle number
current is a topological current,
\begin{equation}\label{current-X}
    J^\mu =\frac{n_0}{6} \epsilon^{\mu\nu\lambda\rho} \epsilon^{abc}
\d_\nu X^a \d_\lambda X^b \d_\rho X^c \, ,
\end{equation}
where $n_0$ is the unperturbed density.
In particular, the particle number density~\cite{Son_2005},
\begin{equation}
J^0 =n_0 \det|\d_i X^a|=\frac{n_0}{6} \epsilon^{ijk} \epsilon^{abc}
\d_i X^a \d_j X^b \d_k X^c \,, 
\end{equation}
is proportional to the Jacobian of the transformation from the external spatial
coordinate $x^a$ to the coordinates frozen within the solid $X^a$.

The particle number current is coupled to the $U(1)$ gauge field $A_\mu$
through a term $A_\mu J^\mu$.  To introduce a nondynamical neutralizing
background with the background charge density $-n_0$, we will turn on a background Kalb-Ramond gauge field
$B_{\mu\nu}=-B_{\nu\mu}$ with a nonzero field strength
\begin{equation}
    H_{ijk}=
    \d_i B_{jk}+ \d_j B_{ki} +\d_k B_{ij} =
\ell^{-3}\epsilon_{ijk}, 
\end{equation}
with the ``magnetic length'' $\ell$ related to $n_0$ by $n_0=\frac1{2\pi}\ell^{-3}$
and couple it to the electromagnetic field $A_\mu$ through a ``BF'' term $-\frac1{2\pi}\epsilon^{\mu\nu\lambda\sigma}B_{\mu\nu}\d_\lambda A_\sigma$.
The Lagrangian for our model can then be written as follows:
\begin{equation}\label{L-ABX}
    \mathcal L =\frac1{6\pi\ell^3} A_\mu \epsilon^{\mu\nu\lambda\rho} \epsilon^{abc}
\d_\nu X^a \d_\lambda X^b \d_\rho X^c  - 
  \frac{B_i B_i}{2e^2}
  -\epsilon(O^{ab}) -\frac{1}{2\pi} \epsilon^{\mu\nu\lambda\sigma} B_{\mu\nu}\d_\lambda A_\sigma \, ,
\end{equation}
where $O^{ab}=\d_i X^a \d_i X^b$, $B^i=\epsilon^{ijk}\d_i A_k$ (and should be distinguished from the components of the Kalb-Ramond two-index tensor $B_{\mu\nu}$), and $\epsilon(O^{ab})$ is the elastic energy of the lattice.  Note that terms involving the time derivatives of $X^a$ or $A_i$ are dropped because we expect the low energy modes will have quadratic dispersion $\omega\sim q^2$.

Introducing the compensating background through the Kalb-Ramond field allows the background to carry a nonzero divergenceless electric current, parameterized through the $B_{0i}$ components.  Indeed, the BF term in the action can be expanded as
\begin{equation}
   -\frac1{2\pi} \epsilon^{\mu\nu\lambda\sigma} B_{\mu\nu}\d_\lambda A_\sigma =- \frac1\pi B_{0i} B^i - \frac{A_0}{2\pi\ell^3}  \, .
\end{equation}
The first term, up to a total derivative, can be written as $J_\text{ext}^i A_i$ where the external current $J_\text{ext}^i=-\frac1\pi \epsilon^{ijk} \d_j B_{0k}$.  This current is divergenceless, consistent with the constant density of the neutralizing background.

Upon varying the Lagrangian with respect to $A_0$, we obtain a constraint:
\begin{equation}\label{VPD}
    \frac{1}{6} \epsilon^{ijk} \epsilon^{abc}
\d_i X^a \d_j X^b \d_k X^c = 1 \, ,
\end{equation}
which means that the transformation from $x^i$ to $X^a$ is volume-preserving.  To the linear order of displacement $\u$, it follows from Eq.~(\ref{VPD}) that $\bnabla \cdot \u = 0$, indicating that the displacement is divergence-free. Here we would like to address the constraint~(\ref{VPD}) at the nonlinear level. One knows that a volume-preserving diffeomorphism (VPD) can be obtained by exponentiating an infinitesimal VPD \footnote{Another way to express the mapping from $x^a$ to $X^a$ is by the Nambu-Poisson bracket. The Nambu-Poisson bracket~\cite{PhysRevD.7.2405,Takhtajan_1994} serves as a generalization of Poisson brackets based on a canonical triple $x^1$, $x^2$, $x^3$, is defined by
\begin{align}
        \{f,g,h\}=\epsilon^{abc}(x)\d_a f(x)\d_b g(x) \d_c h(x) \, .
\end{align}
Equipped with the generalized Poisson bracket, we can represent the VPD constraint in the following form:
\begin{align}\label{VPDshort}
        \{X^a,X^b,X^c\}=\epsilon^{abc} \, .
\end{align}

Then, the Nambu bracket description of the mapping from $x^a$ to $X^a$ is given by
\begin{equation}\label{X-b-bb}
     X^a =x^a-\ell^3\{x^a, x^c, a_c \}+\frac{\ell^6}{2!}\{\{x^a , x^c, a_c\},x^k, a_k\}+\cdots\, .
\end{equation}
The utilization of the Nambu bracket implies that there is no noncommutative field theory description applicable to spatial dimensions higher than two.}; the latter being given by a differential operator $-\xi^i\d_i$ where $\xi^i$ is a divergence-free vector field, $\d_i \xi^i=0$.  We can solve the constraint on $\xi^i$ by introducing a gauge potential $a_i$: $\xi^i=\ell^3\epsilon^{ijk}\d_j a_k \equiv \ell^3 b^i$.  Thus we have
\begin{equation}\label{VPDtransf}
\begin{split}
     X^a &= e^{-\xi^i\d_i} x^a = x^a -\xi^a + \frac12 \xi^k\d_k \xi^a- \frac16 \xi^l \d_l (\xi^k \d_k \xi^a) + \cdots  \\
  & = x^a -\ell^3 b^a + \frac{\ell^6}{2} b^k\d_k b^a + \cdots \,.
\end{split}
\end{equation}
In order to explore the linear regime and find the quadratic Lagrangian, we initiate our analysis by Eq.~(\ref{VPDtransf}). One can then expand the Lagrangian and reach the linearized Lifshitz photon theory in Eq.~(\ref{Lifshitz_photon_theory}). This corresponds to a quadratic dispersion relation $\omega \sim q^2$ which is protected by the ``magnetic" translation symmetry, as described in the subsequent section.

\subsection{Nonlinear multipole symmetries in (3+1) dimensions}
In this Section, our focus is directed toward the exploration of nonlinear dipole and higher multiple symmetries in (3+1) dimensions.

We first consider translation. The operation of translation of the whole lattice by a spatial vector $c^i$ is given by \footnote{An alternative method for determining the nonlinear dipole symmetry can be achieved through the application of the subsequent algebraic approach. Let us introduce the diveregence-free vector
fields $\bm{\xi_1}=\ell^3 \bnabla \times \a'$, $\bm{\xi_2}= \ell^3 \bnabla \times \bar \a$ and $\bm{\xi_3}=\ell^3 \bnabla \times \Tilde{\a}$. The Lie bracket is given by
\begin{equation}
    [\xi^i_1,\xi^i_2]= \{\xi_1, \xi_2\}^i=\xi_1^j\d_j \xi_2^i-\xi_2^j\d_j\xi_1^i=\xi^i_3\, ,
\end{equation}
from which we find
\begin{equation}\label{n-bracket}
    [a'_i, \bar a_j]=-\ell^3 \epsilon_{ijk} (\bnabla \times \a')_i  (\bnabla \times \bar \a)_j \, .
\end{equation}
By employing this algebraic approach in conjunction with Eq.~(\ref{VPDtransf}), one can derive the nonlinear dipole symmetry as expressed in Eq.~(\ref{phi-magn-tr}).}
\begin{equation}
X^a \to X^a(\vec x - \vec c)=e^{-c^i \d_i} X^a\, .
\end{equation}
But since $X^a$ is related to $a_i$ by
Eq.~(\ref{VPDtransf}), one reaches
\begin{equation}\label{transl}
  e^{-\ell^3 (\bnabla \times \a)^i \d_i} \to e^{-c^i \d_i} \ e^{-\ell^3 (\bnabla \times \a)^i \d_i}\, .
\end{equation}
From this we derive the action of translation on the Lifshitz photon $a_i$, it can be expressed as
\begin{equation}\label{phi-magn-tr}
  a_i \to a_i + \frac1{\ell^3}\epsilon_{ijk} c^j x^k -\frac{1}{2} \epsilon_{ijk} c^j (\bnabla \times \a)_k+ \cdots \, .
\end{equation}
This nontrivial form of the transformation law has the consequence that two translations do not commute:
\begin{equation}\label{mag-algebra}
    [P_i, P_j]= i\ell^{-3} \vareps_{ijk} Q_k \, ,
\end{equation}
where under $Q_k$ the Lifshitz photon transforms as $a_k \to a_k + c_k$.  Thus $Q_k$ can be identified with a conserved charge of a $U(1)^{(1)}$ one-form symmetry, which is spontaneously broken, giving rise the Lifshitz photon is a Nambu-Goldstone boson (NGB).

Equation~(\ref{mag-algebra}) should remind one of algebra of the translations on a two-dimensional plane in the presence of a magnetic field---the ``magnetic translations.''  There the commutator of the two magnetic translations is proportional to the $U(1)$ charge, with the coefficient of proportionality being the magnetic field.  

Thus, the Lifshitz photon is a three-dimensional generalization of the Tkachenko mode propagating on a vortex lattice of a rotating two-dimensional superfluid.  As in the latter case,   the quadratic dispersion relation is a consequence of the symmetry.  The Lifshitz photon here is also a NGB ``shared'' between the spontaneously broken $U(1)^{(1)}$ and translational symmetries, with the consequence that  the number of broken generators is not equal to the number of NGBs, which is two in the case of the Lifshitz photon~\cite{NIELSEN1976445,Sch_fer_2001,Watanabe_2013,PhysRevX.4.031057}. 

In Appendix~\ref{generalization} we generalize the Lifshitz photon theory to $d$ spatial dimensions. The algebraic formulation presented in Eq.~(\ref{mag-algebra}) becomes associated with a $(d-2)$-form charge $Q^{(d-2)}$.

The system also realizes rotations as nonlinear higher multipole symmetry, in analogy with the magnetic rotation in the theory of the Tkachenko mode. 
From
\begin{equation}  
  e^{-\ell^3 (\bnabla \times \a)^i \d_i} \to e^{\epsilon^{ijk} \omega_i x_j \d_k} \ e^{-\ell^3 (\bnabla \times \a)^i \d_i}\, ,
\end{equation}
we find that rotation is realized as a higher multipole symmetry
\begin{equation}
    a_i \to a_i + \frac1{\ell^3} \omega_i x^2 +\cdots \, .
\end{equation}

\subsection{Interaction and decay rate of the Lifshitz photon}

The nonlinear nature of the parameterization~(\ref{VPDtransf}) and the dipole symmetry~(\ref{phi-magn-tr}) imply that the effective Lagrangian of the Lifshitz photon must contain nonlinear terms describing its self-interaction.

We will limit ourselves to finding the cubic vertices of interaction.  We first compute the current $J^i$ in Eq.~(\ref{current-X}) by substituting Eq.~(\ref{VPDtransf}),
\begin{equation}\label{J-b-bb}
  J^i = \dot b^i + \frac{\ell^3}2 (\dot b^k \d_k b^i - b^k\d_k \dot b^i) + O(a^3)\, .
\end{equation}
Since $\d_i J^i=0$, one expects that 
one can write $J^i = \epsilon^{ijk}\d_j \Pi_k$.  From Eq.~(\ref{J-b-bb}) one obtaines
\begin{equation}
  \Pi_i = \dot a_i + \frac{\ell^3}2 \epsilon_{ijk} \dot b^j b^k + O(a^3)\, .
\end{equation}
This allows us to rewrite
\begin{equation}
  J^i A_i - \frac{B^2}{2e^2} = B^i \Pi_i - \frac{B^2}{2e^2} + \text{total derivative}\, .
\end{equation}
Integrating over $B_i$ after introducing a Lagrange multiplier enforcing the constraint $\d_i B^i=0$, one then obtains in the action the term
\begin{equation}
  2\pi\alpha \left( e_i - \frac{\ell^3}2\epsilon_{ijk} \dot b^j b^k\right)^2\, ,
\end{equation}
where $\alpha=e^2/(4\pi)$.
The cubic interaction that emerges from this term, after integration by part, can be written as
\begin{equation}\label{cubic1}
  2\pi\alpha \ell^3 e_i e_j \d_i b_j\, .
\end{equation}

Another source of interaction is in the elastic energy.  This includes the quadratic term in the elastic energy, expanded to cubic order in $b$ according to Eq.~(\ref{X-b-bb}), as well as in the cubic term of the elastic energy.  The terms that one obtains are (suppressing spatial indices)
\begin{equation}\label{cubic2}
  (\d b)^3, \quad b d b \d^2 b\, .
\end{equation}

Using the power counting scheme appropriate for the Lifshitz photon theory (in which momentum has dimension 1 and energy has dimension 2), one finds that the coupling constants in front of the terms~(\ref{cubic1}) and (\ref{cubic2}) have dimension $-\frac52$.  Since the decay rate is proportional to the square of the coupling constants, the energy dependence of the decay width of the Lifshitz photon is 
\begin{equation}
    \Gamma(E) \sim  E^{\frac{7}{2}} \, .
\end{equation}
Compared with the Tkachenko mode in two dimensions,
the decay width of the Lifshitz photon tends to zero at a faster rate.

\section{Prospects of higher form symmetry}
\label{sec:prospects}

As we mentioned before, the gauge field $\a$ is the dual gauge field of the physical EM field $\A$. The Wilson loop of $\a$, i.e. $W_{\a} = \exp( \ii \oint_{\cC} d\l \cdot \a)$ is the order parameter of the magnetic 1-form symmetry $U(1)^{(1)}$ of the physical EM field, where $\cC$ is a closed loop in space. And $W_{\a}$ is also the 't Hooft loop of the EM field. Based on the Lagrangian Eq.~\ref{laga}, the scaling dimension of $\a$ is $[\a] = 1/2$, this leads to the observation that the Wilson loop of $\a$ scales as \beqn \langle W_{\a} \rangle \sim \exp(- c L \log L ), \eeqn where $L$ is the perimeter of $\cC$. This is a qualitatively different scaling from that of the 't Hooft loop of an ordinary EM field in the vacuum, which should decay with a perimeter law. A suppressed scaling of the 't Hooft loop of the EM field can usually be attributed to the screening from the electric charges. The strongest screening of the EM field is the condensation of the electric charges, which drives the EM field into a Higgs phase and render the 't Hooft loop decay with an area law. In our case, the EM field is screened by the fluctuation $\u$ of the Wigner crystal of the electric charges, which is a much weaker screening compared with the Higgs machanism, but still leads to a different scaling of the 't Hooft loop. Simple power-counting suggests that, in higher dimensional generalizations of the theory where $\a$ become a $(d - 2)-$form gauge field ($d$ is the spatial dimension rather than space-time dimension), the ``Wilson-membrane" of $\a$ should always violate the perimeter law with an extra logarithmic factor. 

The logarithmic correction to the 't Hooft loop also occurs in another more familiar system: the $(3+1)d$ QED with massless Dirac fermions. First of all, let us assume that there is no Dirac monopoles in the QED, hence there is a strict magnetic $U(1)^{(1)}$ 1-form symmetry. With massive matter fields, the 't Hooft loop should decay with a perimeter law, and the coefficient of the perimeter law is proportional to $1/\alpha$, where $\alpha$ is the fine-structure constant: \beqn \log \langle W_\a \rangle  \sim  - \frac{1}{\alpha} L. \eeqn  However, when the EM field $A_\mu$ is coupled to the massless Dirac fermions, $\alpha$ will be marginally irrelevant in the infrared due to screening from the massless fermions, hence we expect the 't Hooft loop to receive an extra factor of $\log L$ for large $L$: \beqn \log \langle W_\a \rangle  \sim  - \frac{1}{\alpha_0} L \log L. \eeqn This is because the fine-structure constant $\alpha(\mu)$ at energy scale $\mu$ is $\alpha(\mu) \sim \alpha_0/\log(1/\mu)$, and $\mu \sim 1/L$.

\section{Potential connection to other lattice gauge systems}
\label{sec:lattice}

In this section we explore a potential realization of the physics discussed in the previous sections in the context of other systems with a description of lattice gauge theories. One class of such systems is the well-known quantum spin-ice~\cite{Hermele_2004}. The quantum spin ice materials usually have quantum Ising spins from the rare-earth elements that live on the Pyrochlore lattice, which is dual to the diamond lattice in the sense that the sites of the Pyrochlore lattice are the links of the diamond lattice. The largest term of the Hamiltonian is the following:
\begin{equation}
H_0 = \sum_t J_z (\sum_{i=1}^4{\bf S}^z_{t,i})^2 \, .
\end{equation}
The subscript “$t$” labels each tetrahedron of the Pyrochlore lattice; and ${\bf S}^z_{t,i}$ labels the $i$-th spin-1/2 degree of freedom (the $z$-component only) on the tetrahedron. Notice that $H_0$ is simply a nearest neighbor antiferromagnetic interaction on the Pyrochlore lattice. The ground states of $H_0$ consist of all configurations of ${\bf S}^z_{t,i}$ which satisfy $\sum_i {\bf S}^z_{t,i}=0$ for each tetrahedron $t$, i.e. each tetrahedron has two up-spins and two down-spins. 

${\bf S}^z$ is mapped to the (discrete) electric field; and the condition $\sum_i {\bf S}^z_{t,i}=0$ is mapped to the Gauss law constraint $\bnabla
 \cdot \E =0$. Now suppose we turn on spin exchange $\sum_{\<i,j\>} J_{\perp} (S^x_i S^x_j + S^y_i S^y_j)$, at the third order perturbation of $J_\perp / U$, we are going to generate a ring exchange term
\begin{equation}
    H_r \sim \sum_{\hexagon} J_r S_1^+ S_2^- S_3^+ S_4^- S_5^+ S_6^- \, .
\end{equation}
This term will map to the term $H_r \sim J_r \cos(\bnabla \times \A)$, where $J_r \sim J_\perp^3 / J_z^2$.

Now we organize the entire low energy effective Hamiltonian as
\begin{equation}
    H =\frac{U}{2} \E^2 - J_r \cos(\bnabla \times \A)\, .
\end{equation}
Notice that now we allow $\E$ to take all half-integer values, and the first $U$ term will constrain the low energy Hilbert space to $\E = \pm 1/2$.

We can now consider polarizing a fraction of the Ising spin ${\bf S}^z$. If one ${\bf S}^z$ is flipped from ${\bf S}^z = -1/2$ to ${\bf S}^z = +1/2$, it amounts to create an electric field on a site of the Pyrochlore lattice (or the link of the diamond lattice), which violates the constraint $\bnabla \cdot \E =0$ on a pair of nearest neighbor sites of the diamond lattice, i.e. this is equivalent to creating a dipole of gauge charges. More precisely it creates a positive gauge charge on the sublattice A of the diamond lattice, and a negative gauge charge on the sublattice B of the diamond lattice. In general we have the following relation:
\begin{equation}
\begin{split}
        \oint_{\cal A} \E \cdot d\s = n_A - n_B\, ,
        \\
        M \sim \sum n_A + n_B\, .
\end{split}
\end{equation}

\begin{figure}[t]
    \centering
    \subfigure[]{\includegraphics[width=0.36\textwidth]{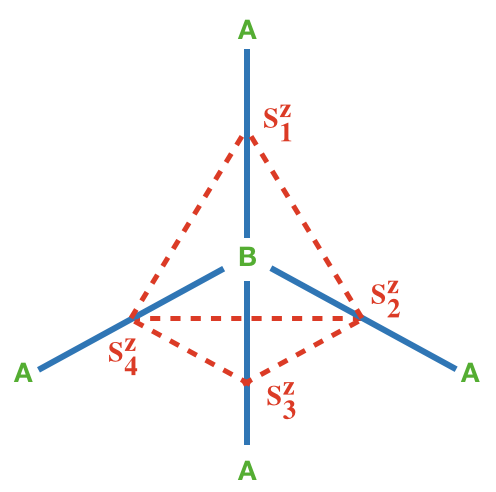}}      
    \qquad
    \subfigure[]{\includegraphics[width=0.26\textwidth]{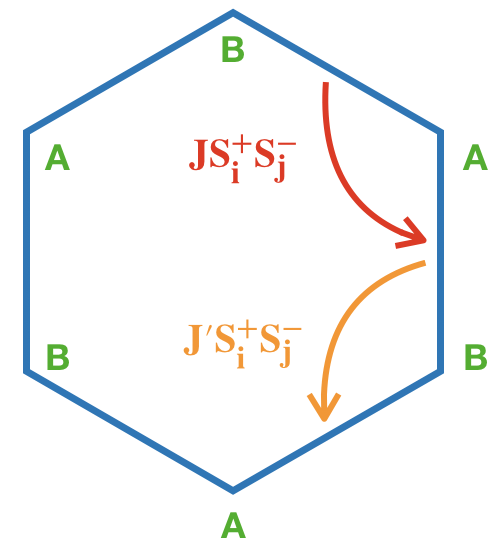}} 
    \caption{(a) shows the basic structure of the Pyrochlore lattice, which is built with corner-sharing tetrahedrons (red); the sites of the tetrahedrons are the links of a dual diamond lattice; the two sublattices of the diamond lattice are labelled A and B. The purpose of (b) is to show that the dynamics of defects (gauge charges) on A and B sublattices can be very different, as long as there is no symmetry that connects A and B sublattices. The dynamics of gauge charges on sublattice A is controlled by spin exchange $J'$ in the sketch, while the dynamics of gauge charges on B is controlled by spin exchange $J$.}
\end{figure}
Here $n_A$ and $n_B$ count the number of ``defects" that violate the original constraint $\sum_i {\bf S}^z_{t,i}=0$ on sublattice $A$ and $B$ of the diamond lattice respectively, and their total number can be controlled by the magnetization $M$ of the system. These defects become mobile with the $J_\perp$ spin exchange term, but $J_{\perp}$ always hops defects within the same sublattice. Defects on sublattice $A$ and $B$ should have the same total number, but they do not necessarily have the same dynamics unless there is a symmetry that connects these two sublattices. Hence it is conceivable that the defects on sublattice $A$ form a Wigner crystal that is incommensurate with the lattice and hence lead to gapless excitations that correspond to $\u$, while the defects on sublattice $B$ are pinned by the lattice or disorder. Then the WC of defects on sublattice $A$ will lead to physics discussed in this paper. We leave more detailed construction in the context of quantum spin ice to future study.

\section{An alternative candidate: ferromagnets}
\label{sec:ferro}

In this Section, we speculate on another possible candidate for a physical system featuring Lifshitz photons: a ``deconfined'' ferromagnet. First we recall some facts about the nonlinear sigma-model of ferromagnets.  The action of the sigma-model is~\cite{10.21468/SciPostPhys.12.2.050,FradkinBook}:
\begin{equation}
    S
  = S_0\! \int\limits_0^1\!d\sigma\! \int\!dt\,d\x\,
  \vareps^{abc} n^a \d_t n^b \d_\sigma n^c
   - \frac J2\! \int\!dt\, d\x\, \delta^{ij} \d_i n^a \d_j n^a \, ,
\end{equation}
where $S_0$ is a parameter related to the volume density of spin.
The model has a 2-form current 
\begin{equation}\label{2-form-current}
    J^{\mu\nu}= \frac1{8\pi} \vareps^{\mu\nu\lambda\rho} \vareps_{abc} n^a \d_\lambda n^b \d_\rho n^c\, ,
\end{equation}
which is by construction conserved, $\d_\mu J^{\mu\nu}=0$.  The location where the 2-form density $J^{0i}$ is nonzero can be identified with the Skyrmions, which are one-dimensional lines (or loops) in three-dimensional space.  

There is an alternative description of the   
$\mathbb{CP}^1$ parametrization of the spin vector: $n^a = z^\+ \sigma^a z$, where 
\begin{equation}
z = \left( \begin{array}{cc} z_1 \\ z_2 \end{array} \right) ,\qquad z^\+ z = 1\, .
\end{equation}
The phase of $z$ is redundant, giving rise to an emergent gauge field $\alpha_\mu$.  The effective action for the ferromagnet can be written as
\begin{equation}
    \mathcal{L}= 2i S_0 z^\+ \d_t z -2 J D_i z^\+  D_i z + \lambda(|z|^2-1)
    \, ,
\end{equation}
where $D_\mu$ is the covariant derivative, $D_\mu z = (\d_\mu -i \alpha_\mu)z$, and $\lambda$ is a Lagrange multiplier enforcing the condition $|z|^2=1$.  The equation of motion for $a_\mu$ can be solved to yield,
\begin{equation}
\alpha_\mu = - \frac i2 \left[ z^\+ (\d_\mu z) - (\d_\mu z^\+) z \right]\, .
\end{equation}
The 2-form current~(\ref{2-form-current}) can be expressed in terms of the photon as
\begin{equation}
    J^{\mu\nu} 
= \frac{1}{2\pi}\epsilon^{\mu\nu\rho\lambda} \d_\rho \alpha_\lambda\, .
\end{equation}
For example, the Skyrmion density vector is 
\begin{equation}
    J_i\equiv J^{0i}= \frac{1}{8\pi} \epsilon^{ijk}  \epsilon_{a b c} n^a  \partial_j n^b \partial_k n^c = \frac{1}{2\pi}\epsilon^{ijk} \d_j \alpha_k\, ,
\end{equation} 
hence a Skyrmion loop is the magnetic flux loop of $\alpha_\mu$.

It is also known~\cite{Papanicolaou:1990sg} that the ferromagnetic nonlinear sigma model possesses nontrivial conservation laws: in addition to the conservation of the one-form charge
\begin{align}
    Q_i = \int\!d\x\ J_i\, .
\end{align} 
all the first moments
\begin{align}
    I_{ij}= \int\!d\x\ x_i J_j\, ,
\end{align}
are also conserved.  In addition, one higher moment,
the angular momentum is expressed by
\begin{equation}
    l_i = \frac{1}{2}\int\!d\x\ x^2 J_i\, ,
\end{equation}
is also conserved. 
The quantities
\begin{equation}\label{cons}
    p_i=\int\!d\x\ \epsilon_{ijk} x_j J_k ,\qquad  l_i= \frac{1}{2}\int\!d\x\ x^2 J_i\, ,
\end{equation}
are similar to the conservation laws that follow from the dipole and the higher multipole symmetries of the nonlinear Lifshitz photon theory.  In particular, there is a nontrivial commutation relation~\cite{Papanicolaou:1990sg}
\begin{equation}\label{nc-momentum}
    [p_x,\ p_y ] \sim i Q_z\, .
\end{equation}

\begin{figure}[b]
\centering
\includegraphics[width=0.8\textwidth]{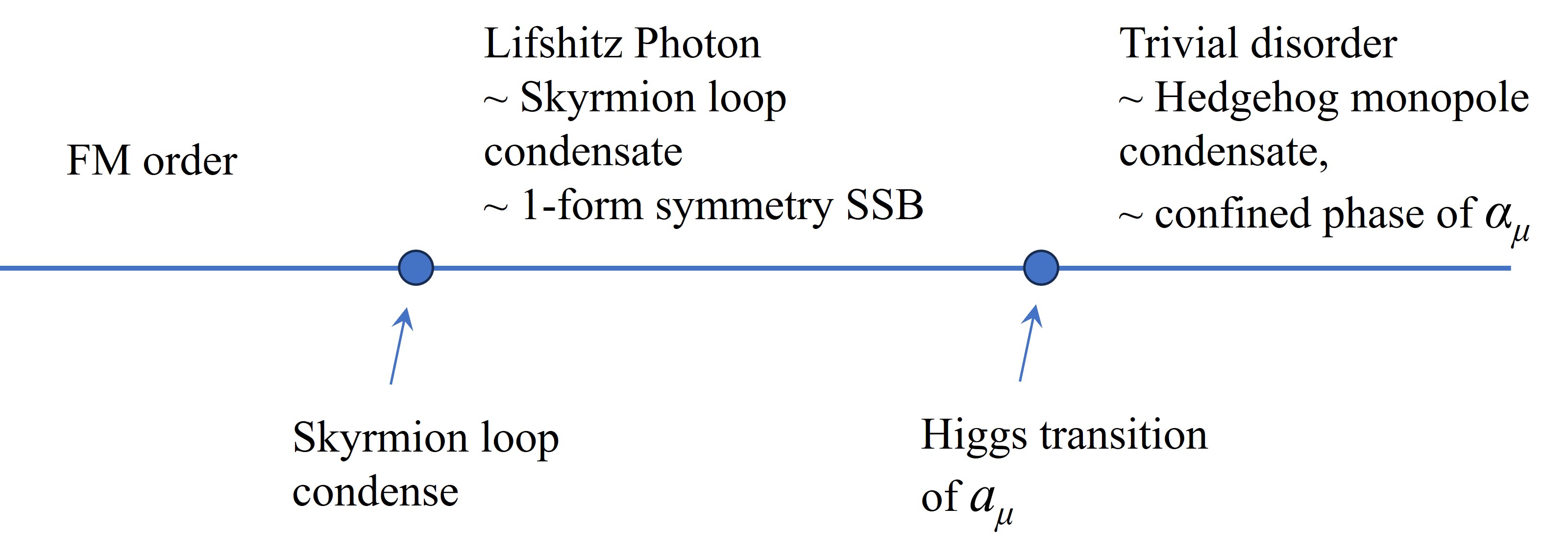}       
\caption{We speculate that starting with a ferromagnet order, one can enter a Lifshitz photon phase by proliferating the Skyrmion loops of the FM order $\vec{n}$. By condensing the hedgehog monopole of the FM order $\vec{n}$ one drives the system into a trivial disordered phase. A hedgehog monopole is the Dirac monopole of gauge field $\alpha_\mu$, and the charge of the Lifshitz photon gauge field $a_\mu$.}\label{phasedia}
\end{figure}

We now speculate that there is a phase 
where Skyrmions proliferate and condense, destroying the ferromagnetic order. The magnetic flux loop of $\alpha_\mu$ is dual to the electric flux loop of $a_\mu$, and in this phase
the gauge field $a_\mu$ becomes the gapless Nambu-Goldstone boson as a result of the condensation of the 1-form symmetry charge, i.e. the electric field flux.  Based on the similarity between the symmetries of the ferromagnetic nonlinear sigma model and that of the nonlinear Lifshitz theory, one can expect that the low energy degree of freedom of such a ``deconfined'' ferromagnet is a Lifshitz photon.  The gauge field describing this Lifshitz photon should be the electromagnetic dual of the gauge field in the $\mathbb{CP}^1$ formulation of the sigma model; its Lagrangian
\begin{equation}
    \mathcal{L} = c_1 e_i e_i - c_2 \left(\d_i b_j + \d_j b_i \right)^2 + g_1 e_i e_j \left( \d_i b_j + \d_j b_i\right) + \cdots \, ,
\end{equation}
should have the nonlinear dipole and higher multipole symmetries. One notices the noncommutative momentum algebra in Eq.~(\ref{nc-momentum}) results in the nonlinear version of the Lifshitz photon theory.  (Note that if one takes into account the the Dzyaloshinskii-Morya interaction~\cite{PhysRevB.92.064412}, then
$\frac{1}{2}\int\!d\x\, x^2 J_i$ is not conserved, which means that the higher multipole symmetry is absent.) 

The Lifshitz photon phase is one type of ``exotic" quantum disordered state of the magnetic system. Of course there could be another completely trivial disordered phase with fully gapped spectrum and no spontaneous symmetry breaking. This trivial disordered phase is allowed in a quantum spin system unless there is a Lieb-Shultz-Matthis theorem that excludes it. The Lifshitz photon phase is actually an intermediate disordered phase sandwiched between the FM order and the trivial disordered phase, and both these disordered phases can be constructed using ingredients of the FM order. The Lifshitz photon phase is driven by proliferating the Skyrmion loops of the FM order $\vec{n}$, while the trivial disordered phase is driven by the condensation of the hedgehog monopole of $\vec{n}$. The hedgehog monopole of $\vec{n}$ is nothing but the Dirac monopole of $\alpha_\mu$, hence in a phase where the Skyrmion loops proliferate while the hedgehog monopoles do not condense, the gauge field $\alpha_\mu$ is in its deconfined Lifshitz photon phase. If we further condense the Dirac monopole of $\alpha_\mu$, the system enters a trivial disordered phase. Notice that $\alpha_\mu$ is the dual gauge field of $a_\mu$, hence the condensation of the Dirac monopole is a Higgs transition of the Lifshitz gauge field $a_\mu$ (Fig.~\ref{phasedia}).

\section{Conclusion}
In this work, we present a nonlinear version of the Lifshitz photon theory applied to diverse condensed matter systems in (3+1) dimensions. Lifshitz photon, which is dual to the physical photon, emerges as a quasiparticle with two transverse polarizations and a
quadratic dispersion relation. Our primary focus lies on the Wigner crystal of charged particles coupled to a dynamical $U(1)$
gauge field, which serves as a prototypical model for our study. We formulate the Lagrangian for the nonlinear Lifshitz photon theory, which is consistent with a global $U(1)$ 1-form symmetry and nonlinearly realized ``magnetic” translation and
rotation symmetries. Moreover, our analysis reveals the energy-dependent decay rate of the Lifshitz photon through the nonlinear theory. Additionally, we explore the generalization of the nonlinear Lifshitz theory to higher dimensions.

In (3+1)-dimensions, we uncover the presence of a 1-form global symmetry. From the dual of the Lifshitz photon, we know that the Wilson loop of $\a$ serves as the order parameter of the magnetic
1-form symmetry $U(1)^{(1)}$ of the electromagnetic field $\A$. Notably, the scaling of the Wilson loop of $\a$ is determined as $\exp(- c L \log L )$. We know the electromagnetic field is screened due to the
fluctuation $\u$ of the Wigner crystal of the electric charges. This screening effect, although weaker than the Higgs mechanism, induces a distinct scaling for the 't Hooft loop. Additionally, when extending the theory to higher dimensions, we observe a deviation of the ``Wilson-membrane" associated with $\a$ from the perimeter law, featuring an additional logarithmic factor.

We investigate the potential realization of other lattice gauge systems. Specifically, we focus on quantum spin ice, where quantum Ising spins live on the Pyrochlore lattice, which is dual to the diamond lattice. By considering two sublattices of the diamond lattice, with defects either pinned by the lattice or subjected to disorder on one of the sublattices. We observe the emergence of intriguing physics discussed in this paper by the Wigner crystal of defects on another sublattice. We defer a more comprehensive construction within the context of quantum spin ice to future investigations. 

Moreover, we explore the “deconfined” ferromagnet as an alternative candidate for establishing connections to the Lifshitz photon theory, employing Skyrmion condensation. The Lifshitz photon phase represents one type of “exotic” quantum disordered state of the magnetic system, serving as an intermediate disordered phase situated between the FM order and the trivial disordered phase. Our study highlights the intriguing possibility of exploring the interconnection between the nonlinear Lifshitz theory~\cite{du2022noncommutative} and systems of quantum Hall ferromagnets~\cite{PhysRevB.47.16419,PhysRevLett.76.2153,Else_2021} or twisted bilayer graphene.  It is
found in the superconductivity mechanism of magic-angle twisted bilayer graphene, while
elementary Skyrmions condensing brings out a superconducting state, which breaks the $U(1)$
symmetry spontaneously~\cite{Khalaf_2021}. Similar facts of the Skyrmion condensation happen in the antiferromagnet system; condensing Skyrmions restores the spin rotational symmetry through a deconfined quantum critical point (DQCP)~\cite{Senthil_2004,PhysRevB.70.144407, PhysRevX.7.031051, XU_2012,senthil2023deconfined}, the Landau-forbidden quantum phase
transitions. We defer these questions to future research. 

\acknowledgments
The authors thank Leon Balents, Matthew Fisher, Tarun Grover, and Eslam Khalaf for helpful discussions.  This work is supported, in part, by the U.S.\ DOE Grant No.\ DE-FG02-13ER41958 and by the Simons Collaboration on Ultra-Quantum Matter, which is a grant from the Simons Foundation (651440, D.T.S.). 
Y.-H.D. would like to thank the Kavli Institute for Theoretical Physics, University of California, Santa
Barbara, supported in part by the National Science Foundation under Grant No. NSF PHY-1748958, the Heising-Simons Foundation, and the Simons Foundation (216179, LB).
\appendix

\section{Generalizations to higher dimensions}\label{generalization}

Now, we aim to extend the nonlinear Lifshitz theory to (d+1)-dimensions with d ($d > 2$) as the spatial dimension. To begin with, the
generalized Lagrangian of the prototypical system~(\ref{L-simplified}) is given by
\begin{equation}
  \mathcal L = - 
  \frac\mu 4
  \left(\d_i u_j + \d_j u_i - \frac2d\delta_{ij}\d_k u_k\right)^2
  - \frac K2 (\d_i u_i)^2
  + e n_0 \u\cdot \E - \frac{\B^2}{2} \, .
\end{equation}
We can then derive the generalization of the linearized Lifshitz theory. To obtain the generalized nonlinear Lifshitz theory, let us introduce the neutralizing background with background density $n_0 = \frac{1}{2\pi} \ell^{-d}$ and include a (d-1)-form field $B_{\mu_1\mu_2\dots\mu_{d-1}}$ with its field strength
\begin{equation}
    H_{\mu_1 \mu_2\cdots \mu_d}=  \d_{[\mu_1}B_{\mu_2\cdots \mu_{d}]} = \ell^{-d}\epsilon_{\mu_1\mu_2\cdots\mu_d}\, .
\end{equation}
The gauge symmetry accompanied by a (d-2)-form gauge parameter $\Lambda_{\mu_1\cdots\mu_{d-2}}$:
\begin{equation}
 \delta_\Lambda B_{\mu_1\mu_2\dots\mu_{d-1}}\equiv \d_{[\mu_1}\Lambda_{\mu_2\cdots\mu_{d-1}]}\, .
\end{equation}

It is straightforward to generalize the particle number current which is a topological current as
\begin{equation}
    J^{\mu_1} =\frac{n_0}{d!} \epsilon^{\mu_1 \cdots \mu_{d+1}} \epsilon^{i_1 \cdots i_d}
\d_{\mu_2} X^{i_1} \cdots \d_{\mu_{d+1}} X^{i_d} \, .
\end{equation}
Henceforth, we can establish the generalized Lagrangian in d spatial dimensions as follows:
\begin{equation}
   \mathcal L =\frac1{d!\pi\ell^d} A_{\mu_1}  \epsilon^{\mu_1 \cdots \mu_{d+1}} \epsilon^{i_1 \cdots i_d}
\d_{\mu_2} X^{i_1} \cdots \d_{\mu_{d+1}} X^{i_d}  - \epsilon(B^{(d-2)}, O^{ab}) -\frac{1}{2\pi} \epsilon^{\mu_1 \cdots \mu_{d+1}} B_{\mu_1 \cdots \mu_{d-1}}\d_{\mu_d} A_{\mu_{d+1}} \, .
\end{equation}
Following the same procedure upon varying the Lagrangian with respect to $A_0$, one finds the VPD constraint \footnote{This constraint can be rewritten as the condition on the generalized Nambu-Poisson bracket of $X^a$,
\begin{equation}
    \{X^{a_1}, X^{a_2}, \cdots, X^{a_d} \}=1\, .
\end{equation}
Recall that the generalized Nambu-Poisson bracket is defined as
\begin{equation}\label{Nambu}
    \{f_1,f_2,\dots,f_d \}=\epsilon^{i_1 i_2\dots i_d}\d_{i_1}f_1\d_{i_2}f_2\dots \d_{i_d}f_d \, .
\end{equation}
}
\begin{equation}
   \frac{1}{d!} \epsilon^{i_1 \cdots i_{d}} \epsilon^{a_1 \cdots a_d}
\d_{i_1} X^{a_1} \cdots \d_{i_{d}} X^{a_d}=1 \, .
\end{equation}
We know the constraint can be obtained by exponentiating an infinitesimal VPD transformation in Eq.~(\ref{VPDtransf}) where $\xi^i$ is the divergenceless vector field $\xi^{i_1} = \ell^d \epsilon^{i_1 \cdot i_d}\d_{i_{2}} \phi_{i_3 \cdots i_d}$. 

\paragraph{Generalized multipole symmetries---}One can find the generalized ``magnetic" translation and rotation symmetries by Eq.~(\ref{VPDtransf}). It is important to note that these symmetries are termed generalized ``magnetic" translations and rotations due to the presence of the d-form field strength $H_{\mu_1 \mu_2\cdots \mu_d}=\ell^{-d}\epsilon_{\mu_1\mu_2\cdots\mu_d}$.

The generalized ``magnetic" translation is the generalized dipole symmetry, can be expressed by
\begin{equation}\label{gen-mag}
  \phi_{i_1 \cdots i_{d-2}} \to \phi_{i_1 \cdots i_{d-2}} + \frac1{\ell^d}\epsilon_{i_1 \cdots i_d} c^{i_{d-1}} x^{i_d} -\frac{1}{2} \epsilon_{i_1 \cdots i_d} c^{i_{d-1}} \epsilon^{i_d \cdot i_{2d-1}}\d_{i_{d+1}} \phi_{i_{d+2} \cdots i_{2d-1}}+ \cdots \, ,
\end{equation}
which satisfies the ``magnetic" translations algebra:
\begin{equation}
    [P_{i_1}, P_{i_2}] = i \ell^{-d} \epsilon_{i_1 \cdots i_d} Q_{i_3 \cdots i_d}\, ,
\end{equation}
with a (d-2)-form charge $Q_{i_3 \cdots i_d}$ of the $U(1)^{(d-2)}$ symmetry. We find the generalized ``magnetic" rotations is the generalized higher multipole symmetry, which is given by
\begin{equation}\label{gen-rot}
  \phi_{i_1 \cdots i_{d-2}} \to \phi_{i_1 \cdots i_{d-2}} + \frac1{\ell^d} \omega^{i_1 \cdots i_{d-2}} x^2 + \cdots \, .
\end{equation}

\paragraph{Generalized Lifshitz theory---} In the linear regime, one gets the divergenceless condition $\nabla \cdot \u=0$ and it can be solved by
\begin{equation}
    u^{i_1}= \ell^d \epsilon^{i_1 \cdot i_d}\d_{i_{2}} \phi_{i_3 \cdots i_d}\, ,
\end{equation}
where $\phi_{i_3 \cdots i_d}$ is the (d-2) higher-form field. Thus, one gets the linearized theory by adding the Lagrange multiplier
\begin{equation}
    \mathcal L= c_1\left(\dot \phi_{i_1 \cdots i_{d-2}} - \d_{i_1} \cdots \d_{i_{d-2}}\phi_0\right)^2 -c_2 \left( \epsilon^{i_1 \cdot i_d}\d_j \d_{i_{2}} \phi_{i_3 \cdots i_d}\right)^2\, .
\end{equation}
The general Lagrangian for the generalized nonlinear Lifshitz theory at higher dimensions that are consistent with $U(1)^{(d-2)}$ symmetry and ``magnetic" translations~(\ref{gen-mag}), contains the combination of the terms
\begin{equation}
  \left(\dot \phi_{i_1 \cdots i_{d-2}} - \d_{i_1} \cdots \d_{i_{d-2}}\phi_0\right)  ,\, \left(\epsilon^{i_1 \cdot i_d}\d_j \d_{i_{2}} \phi_{i_3 \cdots i_d}\right)\, .
\end{equation}
Finally, the energy dependence of the decay rate is 
\begin{equation}
    \Gamma(E) \sim E^{\frac{d+4}{2}}\, .
\end{equation}

\bibliography{Lifshitz}{}

\end{document}